\documentclass[letterpaper, 10 pt, conference]{ieeeconf}  

\IEEEoverridecommandlockouts                             
\usepackage{tikz}
\usepackage{dsfont}

\usepackage{amsthm}
\usepackage{amssymb}
\usepackage{romannum}
\usepackage{xcolor}
\usepackage{cite}
\usepackage{amsmath,amssymb,amsfonts}
\usepackage{textcomp}
\usepackage{graphicx} 
\usepackage{algorithm}
\usepackage[capitalize]{cleveref}
\usepackage{algpseudocode}
\usepackage{url}
\usepackage{bbm}

\renewcommand{\baselinestretch}{0.93}
\usepackage[inkscape=false]{svg}
\usepackage{pgfplots}
\usepackage{pifont}
\pgfplotsset{compat=1.8}
\usepackage{textcomp}
\theoremstyle{definition}

\newtheorem*{assumption*}{Assumption}
\newtheorem{theorem}{Theorem}

\newtheorem{lemma}[theorem]{Lemma}

\newtheorem{definition}{Definition}
\theoremstyle{remark}

\usepackage{multirow}
\usepackage{tikz}
\usepackage{overpic}
\usepackage{caption}
\title{\LARGE \bf Iterative VCG-based Mechanism Fosters Cooperation \\in Multi-Regional Network Design
}

\newcommand{\parens}[1]{\left( #1 \right)}

\author{Mingjia He$^{1}$, Yannik Werner$^{1}$, Andrea Censi$^{1}$, Emilio Frazzoli$^{1}$, and Gioele Zardini$^{2}$%
\thanks{This work was supported by the ETH Zürich Mobility Initiative (MI-03-22) and ETH Zürich Foundation (2022-HS-213).}%
\thanks{$^{1}$Institute for Dynamic Systems and Control, ETH Zürich, 8092 Zürich, ZH, Switzerland 
(e-mail: minghe@ethz.ch; ywerner@student.ethz.ch; acensi@ethz.ch; emilio.frazzoli@idsc.mavt.ethz.ch).}%
\thanks{$^{2}$Laboratory for Information and Decision Systems, Massachusetts Institute of Technology, Cambridge, MA, USA. 
(e-mail: gzardini@mit.edu).}
}
\begin{document}

\maketitle
\thispagestyle{empty}
\pagestyle{empty}

\begin{abstract}

Transportation network design often involves multiple stakeholders with diverse priorities. 
We consider a system with a hierarchical multi-agent structure, featuring self-optimized subnetwork operators at the lower level and a central organization at the upper level.
Independent regional planning can lead to inefficiencies due to the lack of coordination, hindering interregional travel and cross-border infrastructure development, while centralized methods may struggle to align local interests and can be impractical to implement.
To support decision making for such a system, we introduce an iterative VCG-based mechanism for multi-regional network design that fosters cooperation among subnetwork operators. 
By leveraging the Vickery-Clarke-Groves (VCG) mechanism, the framework determines collective investment decisions and the necessary payments from both operators and the central organization to achieve efficient outcomes. 
A case study on the European Railway System validates the effectiveness of the proposed method, demonstrating significant improvements in overall network performance through enhanced cross-region cooperation.
\end{abstract}

\section{INTRODUCTION}
\noindent Rail infrastructure serves as a cornerstone of transportation systems, enabling efficient, low-carbon passenger and freight mobility while driving economic growth and supporting global sustainability.
As the most emissions-efficient transport mode, rail accounts for only 1\% of global transport emissions while facilitating 7\% of passenger travel and 6\% of freight transport~\cite{IEA2023}. 
The global rail infrastructure market, valued at \$51.5 billion in 2024, is projected to expand to \$71.01 billion by 2030, with an annual growth rate of 5.5\%~\cite{railmarket2024}.
However, despite this growth, rail transport has substantial  potential to increase its market share relative to air and road transportation.
Eurostat data from 2022~\cite{eurostat} highlights this potential, noting Switzerland's leading rail share of 17\% among EU and EFTA countries, commpared to 65\% for road transport. 
Austria follows with a 12\% rail share, which is significantly lower compared to air travel (by 13\%) and road transport (by 49\%).

\noindent A key challenge in rail transport is the lack of coordination between regions~\cite{cats2025long}. 
Due to the interconnected nature of railway systems, inefficiencies in one region can negatively impact overall network performance.
Addressing this challenge requires robust central authorities, such as governments and stakeholder associations, which play essential regulatory and financial roles.
These organizations spearhead initiatives and provide financial assistance to strengthen regional and national infrastructure.
For example, the European Commission introduced the Trans-European Transport Network (TEN-T)~\cite{TenT}, a policy to guide the planning of the EU's transport network and establish infrastructure requirements in multiple modes of transportation, including railways, waterways, roads and airports.
Similar strategic efforts include China's Belt and Road Initiative (BRI)~\cite{bri}, the Infrastructure Investment and Jobs Act (IIJA)~\cite{iija}, and the Trans-Asian Railway (TAR) Network~\cite{tarn}.

\begin{figure}[tb]
    \centering
    \includegraphics[width=1\linewidth]{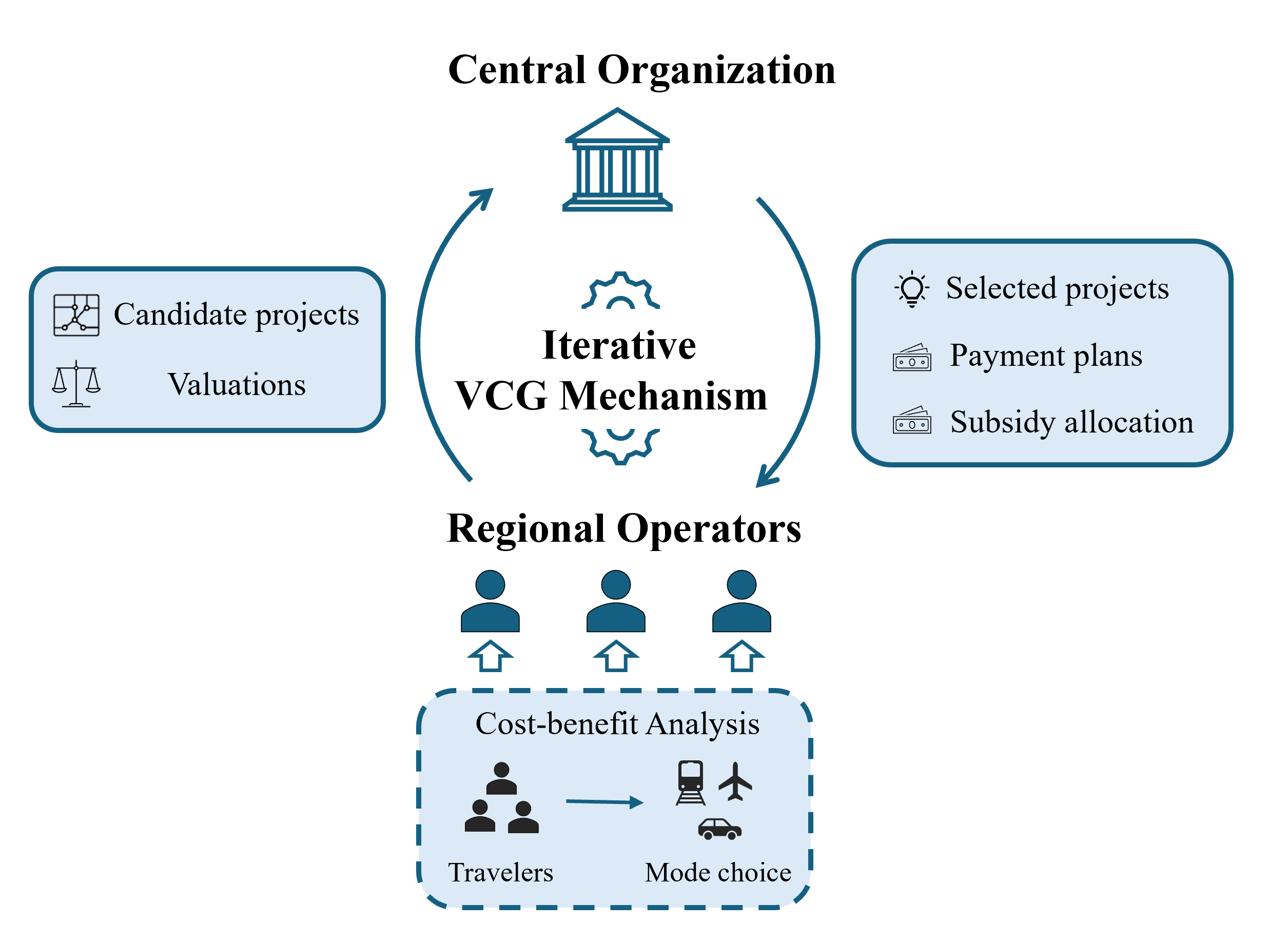}
    \caption{\small In the iterative VCG-based mechanism, the central organization selects infrastructure projects, determines payment plans, and allocates subsidies to foster cross-region cooperation. Self-interested regional operators evaluate projects based on local social welfare. This iterative process supports long-term infrastructure planning.}
    \label{fig:intro}
\end{figure}

\noindent In this work, we consider a multi-regional network design problem characterized by a hierarchical stakeholder structure, consisting of regional operators at the lower level and a central organization at the upper level.
Regional stakeholders, such as local governments or transport agencies, allocate resources based on regional priorities, whereas the central organization strategically subsidizes infrastructure projects to incentivize cooperative investments aimed at maximizing overall social welfare.
This arrangement facilitates regional collaboration by attracting both local and external investment, creating opportunities for stronger interregional partnerships.
In cases where collective regional investment falls short of meeting project costs, central government subsidies become instrumental in enabling critical infrastructure projects and promoting regional cooperation.
Key considerations in this multi-level framework include determining the specific infrastructure projects to undertake, defining the financial contributions of each regional operator, and allocating central subsidies effectively.

\noindent To address these issues, we propose an iterative Vickrey-Clarke-Groves (VCG)-based resource allocation framework (illustrated in \cref{fig:intro}) as a decision-making tool.
The VCG mechanism, known for its auction-based incentive compatibility in game theory, is widely employed in sectors such as energy markets and digital advertising~\cite{karaca2019designing, deng2021towards}. 
In our framework, the central organization acts as the coordinating entity, with regional stakeholders participating as bidders, thereby enabling efficient, transparent, and cooperative resource allocation.
We extend the traditional VCG mechanism\cite{green_incentives_1978} by integrating the central organization's willingness to invest and incorporating budget constraints of regional operators.
Payment responsibilities of operators are determined by assessing the pivotal role of their participation.
Furthermore, the central organization identifies infrastructure projects lacking adequate regional investment despite fair contributions.
Finally, we evaluate the effectiveness of our approach through numerical experiments based on the European Rail Network.

\subsection{Related Research}
\noindent 
Transportation systems involve multiple interdependent subsystems with self-interested decision-makers. Designing each subsystem independently often leads to suboptimal outcomes due to a lack of coordination and the neglect of the broader impact on the entire system~\cite{cats2025long}. 
Recent studies explored integrated design approaches by expanding the geographical scope and incorporating multiple transportation modes to optimize overall network performance. 
Grolle et al.~\cite{grolle2024service} investigated the centralized design of the European high-speed rail network. The results revealed strong network integration through key corridors linking multiple countries, highlighting the importance of cross-border cooperation.
Luo et al.~\cite{luo2021multimodal} developed a joint pricing and network design problem for multimodal mobility, aiming to enhance the seamless integration of conventional transit services with emerging Mobility-on-Demand (MoD) services.
In recent research, game-theory-based methods have gained increasing attention due to their ability to account for individual rationality in network design problems.
He et al.~\cite{he_co-investment_2024} proposed a co-investment and payoff-sharing mechanism for the multi-regional network design problem, aiming to align the interests of local decision makers and ensure a fair distribution of the overall payoff.
 
\noindent Mobility stakeholders operate within a hierarchical structure, where the central organization can influence operators' decision making through regulations, policies, and mechanism design. Zardini et al.~\cite{zardini2021game,zardini2023strategic} applied a Stackelberg game framework to model interactions between municipalities and mobility operators, demonstrating how tax policies influence operator strategies and shape future mobility scenarios. 
Nie et al.~\cite{nie2022strategies} examined the electrification process of the bus industry using evolutionary game theory, and analyzed how carbon trading subsidies influence bus operators' decisions regarding the purchase of electrified buses.

\noindent 
In the VCG mechanism, the central organization employs the selection and payment rules to allocate resources among participants.
It has been implemented across various engineering fields, and the properties are analyzed in specific application cases due to its desired theoretical properties
\cite{sessa2017exploring,karaca2019designing, ding2023mechanism, le2018pareto}.
Phuong~\cite{le2018pareto} introduced budget constraints by using a truncation VCG mechanism and shows Pareto optimality and generic incentive compatibility for multi-good auctions in the case of single-minded participants.
Karaca et al.~\cite{karaca2019designing} focused on reverse auctions for continuous goods and used coalitional game theory to investigate the conditions required for coalition-proof outcomes. They applied this framework to the 14-Bus Test System and the Swiss reserve procurement auction based on data from electricity markets.
Ren et al.~\cite{ren2024truthful} applied the VCG mechanism to the task offloading problem in vehicular networks, modeling the trading relationships between vehicles and edge nodes to determine optimal task offloading and pricing strategies.
For the multi-regional network design problem, the central organization can enhance collaboration among local operators by organizing the selection of projects for joint investment and efficiently allocating subsidies. 
The VCG mechanism offers an effective approach to resource allocation by encouraging local operators to reveal their true valuations to a central authority. 
In addition, according to the Clarke pivot rule, each operator compensates for the external impact of their participation (see Section \ref{subsec:m_vcg}), making the mechanism suitable for infrastructure planning scenarios that involve multiple stakeholders;
However, there are still gaps in adapting this mechanism to real-world multi-regional network design problems.
Specifically, it remains unclear how to incorporate the budget constraints of both operators and the central organization, as well as how to account for the central organization’s preferences in project selection.

\subsection{Statement of Contribution}
\noindent The contributions of this work include \textnormal{i)} developing an iterative VCG-based resource allocation framework for the multi-regional network design problem, enabling the central government to select public projects and allocate payments from both operators and the organization itself. Additionally, we \textnormal{ii)} analyze the theoretical properties of the proposed framework with budget constraints and the central organization’s preferences, and \textnormal{iii)} conduct a European railway case study to validate its effectiveness.

\noindent 
The paper is organized as follows. 
The key concepts in the multi-regional network design problem are introduced in Section \ref{sec:Multi-Regional} 
and an iterative VCG-based investment allocation framework is proposed in Section \ref{sec:vcg}. Numerical experiments are conducted in Section \ref{sec:exp} and conclusions are drawn in Section \ref{sec:conclusion}.

\section{Multi-Regional Mobility System}
\label{sec:Multi-Regional}
\subsection{Mobility Network}
\noindent We model a mobility network involving multiple countries, each managed by a local operator~$i \in \mathcal{I}=\{1,2,\dots, I\}$ and multiple transportation modes $m \in \mathcal{M}=\{R, A, C\}$.  Specifically, we focus on rail, airplane, and car in this study.
The network can be modeled as a labeled undirected graph  $\mathcal{G} =(\mathcal{U}, \mathcal{E}, \mathcal{L}_u, \mathcal{L}_e)$, where $\mathcal{U}$ is the set of vertices  $u \in \mathcal{U}$ representing cities, $\mathcal{E} \subseteq \mathcal{U} \times \mathcal{U}$ is the set of edges between cities representing the transportation links, $\mathcal{L}_u: \mathcal{U} \rightarrow \mathcal{Z}_u$ is the labeling function from the set of vertices onto the set of vertex labels $\mathcal{Z}_u$, and $\mathcal{L}_e : \mathcal{E} \rightarrow \mathcal{Z}_e$ is the labeling function from the set of edges onto the set of edge labels. 
Specifically,
a vertex label~$z_u \in \mathcal{Z}_u$ is defined as $z_u = (r_u, d^m_u, s_u) \in \mathcal{Z}_u = \mathcal{I} \times  \mathbb{R}^3 \times \mathbb{R}_{+}$, where 
$r_u$ denotes the country to which the city belongs,
$d^m_u$ represents access/egress travel distance for mode $m$, i.e., travel distance between the city center and the nearest service stations, which depends on the transportation mode (i.e., an airport for air travel, a train station for rail, highway entrances for car).
If there are no service stations for mode $m$ in the city, then $d_u^{m}$ is set to infinity.
Furthermore,~$s_u$ represents the population of the city.
An edge label $z_e \in \mathcal{Z}_e$ is defined by $z_e = (i_e, t^\mathrm{con}_e, l_e, v_e, c_e) \in \mathcal{Z}_e = \{0,1,2\} \times \mathbb{N}_{+} \times \mathbb{R}_{+} \times \mathbb{R}_{+}$, 
where $i_e = 0$ means the edge is not implemented and can not provide service, $i_e= 1$ means the edge is implemented and $i_e = 2$ means under construction. 
$t^\mathrm{con}_e$ is the time horizon remaining for construction, 
$v_e$ is the travel speed, 
$l_e$ is the edge length,
and $c_e$ is the implementation cost. 
The edge set of transport mode $m$ is denoted as $(\mathcal{E}_m)_{m \in \mathcal{M}} \subseteq \mathcal{U} \times \mathcal{U}$, where 
$
\bigcup_{m \in \mathcal{M}} \mathcal{E}_{m} =\mathcal{E}
$.
The railway network can be divided among countries. 
The set of vertices is divided into disjoint subsets $\mathcal{U}^i$, ensuring $ \bigcup_{i \in \mathcal{I}} \mathcal{U}^i = \mathcal{U}$. Rail edges consist of intra-regional edges $ \mathcal{E}^i_{R} = \{(o,d) \mid o,d \in \mathcal{U}^i\}$ and cross-border edges $ \mathcal{E}^{ij}_{R} = \{(o,d) \mid o \in \mathcal{U}^i, d \in \mathcal{U}^j, i \neq j\}$, forming the complete rail graph
$
\bigcup_{i \in \mathcal{I}} \mathcal{E}^i_{R} \cup \bigcup_{i,j \in \mathcal{I}, i \neq j} \mathcal{E}^{ij}_{R}
=\mathcal{E}_{R}
$.
\begin{figure}[tb]
    \centering
    \includegraphics[width=1\linewidth]{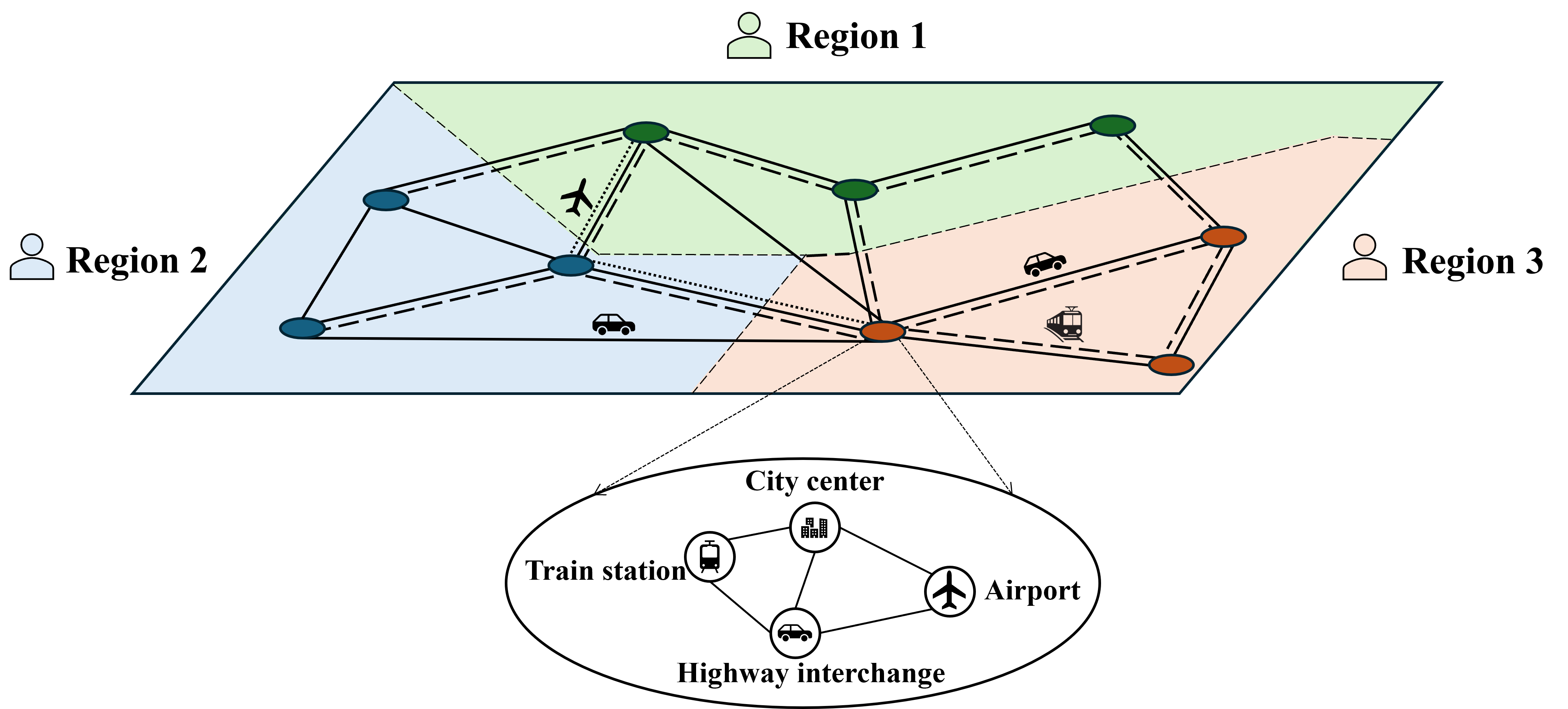}
    \caption{Multi-regional multi-modal mobility network.}
    \label{fig:net}
\end{figure}
\subsection{Travel Demand}
\noindent We consider the intercity trip request~$r_{ij}=(i,j,q_{ij}) \in \mathcal{R} = \mathcal{U} \times \mathcal{U} \times \mathbb{R}_{\geq 0}$, where $i,j$ are the origin and destination cities ($i \neq j$), respectively, and $q_{ij}$ represents the total number of travel requests per year.
Let $\mathcal{P}^m_{ij}$ and $\hat{P}^{m}_{ij}$ denote
the set of paths and the shortest path from city $i$ to city $j$ for mode $m$, respectively. Plane paths may only include direct flights and dual-trips between a car drive and a flight.
The travel distance $d^{m}_{ij}$ when using mode $m$ between cities $i$ and $j$ can be calculated by:
\begin{align}
    d^{m}_{ij}&= 
    \begin{cases}
    \operatorname{min}_{\hat{P}^{m}_{ij} \in \mathcal{P}^m_{ij}}  \sum_{e \in \hat{P}^{m}_{ij}} l_e, \quad &\text{if } \mathcal{P}^m_{ij} \neq \emptyset,\\
        \infty, &\text{otherwise.}
    \end{cases} 
\end{align}
The travel time for the intercity trips $t^m_{ij} \in \mathbb{R}_{\geq 0}$ includes the components: access and egress time to the city center by car using an urban driving speed $v^{u}$, the travel time between cities based on the travel distance $d^m_{ij}$ and mode-specific incurred waiting time along the way $t^m_\mathrm{w}$. The travel time is transformed into monetary value based on component-specific Value-of-Time parameters, $\gamma_\mathrm{a},\gamma_\mathrm{w}, \gamma^\mathrm{m}_t$ \cite{air_passenger_vot}, \cite{VOTEurope_car_train}.
\begin{align}
    c^m_{ij} &= \gamma_\mathrm{a} \cdot \frac{d_{i}^m + d_{j}^m}{v^{u}}+ \gamma_\mathrm{w} \cdot t^\mathrm{m}_\mathrm{w} + \gamma^\mathrm{m}_t \cdot \frac{d^m_{ij}}{v^{m}},
\end{align} 
The proportion of trips $p^m_{ij} \in [0,1]$ served by transport mode $m$ can be calculated via:
\begin{align}
    p^m_{ij} = \frac{e^{-\beta_\mathrm{c} \cdot c^m_{ij}}}{\sum_{m \in \mathcal{M}} e^{- \beta_{c}  \cdot c^m_{ij}} }.
\end{align}
where $\beta_{c}$ corresponds to the sensitivity of travel cost on mode selection\cite{transport_cost_beta}.
The passenger's travel decisions are predicted solely on travel time, but arbitrary extensions beyond the most decisive variable travel time, such as comfort and safety, are possible\cite{small_VoT}.
The travel demand model is responsive to changes in network design, meaning that modifications to the rail network can impact travel costs and, in turn, influence travelers' choice of transportation modes.

\subsection{Mobility Stakeholders}
\noindent Mobility stakeholders include the regional operators $ i \in \mathcal{I} $, and a central organization, denoted by $ i = 0 $.
Each stakeholder $ i \in \mathcal{I} \cup\{0\} $ is characterized by a tuple $(B_i, u_i)$, where $B_i \in \mathbb{R}_{+} $ represents the budget, and $ u_i: \mathcal{X} \to \mathbb{R} $ is a quasilinear utility function, mapping from the project set $\mathcal{X}$ to a utility value:
\begin{align}
u_i(x) = b_i(x) - p^x_i,
\end{align}
where $x \in \mathcal{X}= \{x_0, x_1, \dots, x_K\}$ represent an infrastructure project, $ p^x_i \in \mathbb{R}_{+}$ denotes the investment required for project $x$, 
and $b_i: \mathcal{X} \rightarrow \mathbb{R}$ is the true benefit function of operator $i$, which is private knowledge. 
\begin{align}
    & b_i(x) =  \mathrm{Rev}_i
    \left(
    \mathcal{E}_R^\mathrm{imp} \cup \{x\}
    \right) - \mathrm{Rev}_i
    \left(
    \mathcal{E}_R^\mathrm{imp}
    \right),\\
    & \mathrm{Rev}_i
    \left(\mathcal{E}^\mathrm{imp}_R\right) = \sum_{j \in \mathcal{I}}\sum_{k \in \mathcal{I}} \tau^i_{jk}d^R_{jk}\gamma^R_{p}D^R_{jk}.
\end{align}
The true benefit of a project is the additional revenue it generates. Operators determine this by comparing the revenue from the current rail network, $\mathcal{E}_R^\mathrm{imp}$, to the revenue after including the new edge $x$, i.e., $\mathcal{E}_R^\mathrm{imp} \cup {x}$.
Let $\tau^i_{jk}$ denote the proportion of travel distance in city $i$ along the shortest rail route between cities $j$ and $k$. The revenue is calculated based on the travel distance $\tau^i_{jk} d^R_{jk}$, the service price $\gamma^R_{p}$, and the rail travel demand, given by $D^R_{jk} = D_{jk} p^R_{jk}$.
We assume that operators can invest in projects across the entire network, as the benefit of the operator $i$ from local railway network $G^R_i$ is interdependent with the rest of the network $G^R_j (i \neq j)$, due to cross-border travel requests.
It is noted that the construction of a new link may not have positive impacts on all the local benefits, as it depends on the shortest paths and travelers' mode choice. 
The central organization's utility function is assumed to align with social welfare, i.e., $u_0 = W(X_s)$. The \emph{social welfare function} $W: \mathcal{X}_s \to \mathbb{R}_{+}$ maps the selected project set $\mathcal{X}_s$ to social welfare, which is defined as the sum of the true benefits of the operators:
\begin{align}
W(X_s) = \sum_{i \in \mathcal{I}} b_i(X_s).
\end{align}
Considering the investment strategies of the central organization, a common approach is to allocate resources primarily to projects offering the highest social welfare.
However, this method tends to overlook the value of operator's contributions, especially their potential to foster interregional collaboration.
In this work, we investigate strategies for leveraging the central organization's investment resources to actively improve cooperation among regional operators.
Specifically, we examine scenarios where collective contributions from operators fall short of project costs, indicating that regional collaboration alone is insufficient.
In such cases, targeted financial incentives from the central organization become necessary to bridge funding gaps, ensuring critical infrastructure projects can be realized effectively.

\section{Iterative VCG-based Network Design} 
\label{sec:vcg}
\noindent The mobility system consists of self-interested local operators, each responsible for managing individual regions, and a central organization that aims to maximize social welfare. 
To facilitate decision-making in the network design process, we propose an iterative VCG-based framework in this section.

\subsection{Model Overview}
\noindent Let $M=(\mathcal{T}, \mathcal{X},V, F, P)$ denote the iterative VCG-based mechanism (IVCG), 
where the planning horizon is indexed by $t \in \mathcal{T} = \{0,1,2, \dots, T\}$, and $\mathcal{X}$ is the set of candidate projects.
Let $V=\left( v_i \right)_{i \in \mathcal{I}}$ denote the valuation functions for all operators, where each operator $i$ has a private valuation function $v_i : \mathcal{X} \rightarrow \mathbb{R}_{\geq 0}$. 
For each year $t$, the operators report their valuations of projects to the central organization.
These reported valuations reflect the estimated benefits or local social welfare gains that would result from implementing each project.
Based on the submitted valuations, the central organization then selects projects for joint investment using the project selection function $F: \mathcal{V} \rightarrow \mathcal{X}$ (see \cref{subsubsec:ps}).
The subsidized payment function $P: \mathcal{V} \times \mathcal{X} \rightarrow \mathbb{R}^{I+1}$ maps from the valuation set to the payment set to determine the payments for both the operators and the central organization (see \cref{subsubsec:spf}).  
The mechanism runs in multiple rounds each year, selecting one project for implementation per round.
The process continues until either no beneficial projects remain or the central organization uses up its budget. 
After each year's implementation, local operators can allocate their remaining budget to local projects.


\subsubsection{Variables}
The decision variables are $(X^t, p^t, p^t_0, h^{t}_{i}) \in \mathcal{X} \times \mathbb{R}_{\geq 0}^\mathcal{I} \times \mathbb{R} \times \mathbb{R}_{\geq 0}^\mathcal{I}$, where $X^t=(X_s^t, X_l^t)$ represents the joint invested projects and local designed projects.
Furthermore, $p^t = (p_i^t)_{i \in I}$ denotes the payments from operators to the selected projects, $p^t_0$ represents the subsidy provided by the central organization, and $h^{t}_{i}$ is the budget used in local design in year $t$.
The system state is given by $(\mathcal{E}^{t,\mathrm{imp}}_R, B^t, B^t_0) \in \mathcal{E}_R \times \mathbb{R}_{\geq 0}^I \times \mathbb{R}_{\geq 0}$, where $\mathcal{E}^{t,\mathrm{imp}}_R$ denotes the implemented railway edges, $B^t = (B^t_i)_{i \in \mathcal{I}}$ represents the available budget for operators at year $t$, and $B^t_0$ is the available budget of the central organization.
The state variables are updated according to the following relations, where for all $i \in \mathcal{I}$ and $t \in \mathcal{T} \setminus \{0\}$:
\begin{align}
    &B^{t}_i = \delta B^{t-1}_i - p^{t}_i- h^{t}_{i} + a^{t}_i ,\\
    &B^{t}_0 =  \delta B^{t-1}_0 - p^{t}_0 + a^{t}_0, \\
    &\mathcal{E}^{t,\mathrm{imp}}_R=\mathcal{E}^{t-1,\mathrm{imp}}_R \cup X^t.
\end{align}
For each year, the available budget is updated with a budget increase $a^{t}_i$ and a discount factor $\delta$, according to the inflation rate.
This iterative process is repeated until the mechanism selects the null project or the central organization runs out of budget, i.e., no project is profitable with the available funds. 
\subsubsection{Performance Metrics}
To evaluate the performance of the system, we define three metrics:
\textit{Local Benefit} assesses the benefit individual operators receive from the mechanism.
\textit{System Social Welfare} measures the total social welfare derived from all selected projects; 
\textit{Subsidy Efficiency} represents how effectively the subsidies are used across all planning horizons.
\begin{itemize}
    \item Local Benefit: $\mathrm{LB}=\sum_{t}b_i(X^t_s)$,
    \vspace{0.2em}
    \item System Social Welfare: $\mathrm{SSW}=\sum_{t}W(X^t_s)$,
    \vspace{0.2em}
    \item Subsidy Efficiency: $\mathrm{SE}=\frac{\sum_{t}W(X^t_s)}{\sum_{t}p^t_0}$.
\end{itemize}
\subsection{VCG-based Investment Allocation with Subsidy} \label{subsec:m_vcg}
\noindent For every year of investment, the modified VCG mechanism $M'=( \mathcal{X},V, F, P)$  is applied to determine the invested projects and the distribution of payments.
For clarity, the horizon index $t$ is not included in this demonstration.
\subsubsection{Project Selection for Social Welfare Maximization} \label{subsubsec:ps}
The goal of project selection is to choose the project that maximizes social welfare. 
The set of candidate projects, denoted by $\bar{X}$, must meet the condition that the total reported valuations are greater than the project cost. 
The project selection rule $F=f(v)$ selects the project with the highest total valuation from all operators:
\begin{align}
    \bar{X} &= \{x \in \mathcal{X} | \sum_{i \in \mathcal{I}} v_{i}(x) > c_x \} \label{eq:allowed_investments}, \\
    f(v) &= \begin{cases}
        \underset{x \in \bar{X}}{\operatorname{argmax}} \sum_{i \in  \mathcal{I}} v_{i}(x), \quad & \text{if } \bar{X} \neq \emptyset, \\
        0, & \text{otherwise}.
    \end{cases} \label{eq:social_choice}
\end{align}

\subsubsection{Subsidized Payment Function}\label{subsubsec:spf}
The modified VCG mechanism determines the payments made by both the participants and the central organization.
For the payment distribution among operators, we adopt the concept of participation externality, which differs from the following two traditional perspectives:  1) the project can only be invested by local operators, or 2) the payment is based solely on the operator’s own benefit.
We argue that if the presence of the operator $i$ changes the project chosen for collective investment, the operator should compensate for the welfare loss experienced by other regions. 
Based on this consideration, the payment function for operator $i \in \mathcal{I}$ with reported valuations $v \in \mathcal{V}$ are given according to the Clarke Pivot Rule\cite{green_incentives_1978}:
\begin{align}
    p_i\parens{v,f(v)} = \begin{cases}
        \underset{j \neq i}{\sum} v_{j}\parens{f(v_{-i})} - \underset{j \neq i}{\sum} v_{j}\parens{f(v)}
        \quad & \text{if } \bar{X} \neq \emptyset, \\
        0, & \text{otherwise}.
    \end{cases}
    \label{eq:payments1-n}
\end{align}
Specifically, the payment to the operator $i$ can be interpreted as the difference between the total valuations of all other operators when operator $i$ is excluded ($\sum_{j \neq i} v_{j}(f(v_{-i}))$) and the total valuations of all other operators when operator $i$ is included ($\sum_{j \neq i} v_{j}(f(v))$).

\noindent However, the payments from participants may not always be sufficient to cover the full construction cost of the selected project ($c_{f(v)}$). 
In such cases, an additional payment from the central organization is necessary to ensure the implementation of the chosen project.
We utilize the resources of the central organization to bridge the gap between the total cost and the operators' contributions. The payment from the central organization, denoted by $p_0$, can be determined by:
\begin{align}
    p_0\left(v,f(v)\right) 
    = \max\left(
    c_{f(v)}-\sum_{i \in \mathcal{I}} p_i\left(v,f(v)\right), 0
    \right).
    \label{eq:payments0}
\end{align}


\subsubsection{Admissibility Check (AC)}
We note that the payment rule in VCG is not designed to benefit the central organization, as they may end up paying a substantial amount depending on the project's cost and the total payments collected\cite{cramton_lovely_2005}. 
In practice, however, the central organization has its own willingness to pay. 
We then introduce the investment ratio $\alpha \in [0,1]$ for the project admissibility check.
The decision to implement a project will only be admissible if the required subsidy does not exceed a specified $\alpha$ of the project's cost (in \cref{eq:adm}):
\begin{align}
    p_0 \left(v,f(v)\right)  &\leq \alpha c_{f(v)}.
    \label{eq:adm} 
\end{align}
If this condition is not met, the project $f(v)$ will be excluded from this cycle, and the selection process will continue.

\noindent As operators have limited budgets, projects and payments can be infeasible if an operator is unable to pay their share. Considering the budget constraints, let $\eta_i$ denote the available budget of the operator $i$. 
We then introduce admissible preference profiles, which ensure that operators can afford their contribution irrespective of other operators' valuations.

\begin{definition}
    [Admissible Preference Profiles]
    A preference profile  $v_i \in V_i$  is \emph{admissible} if for all  $v_{-i} \in V_{-i}$  with  $v = (v_i, v_{-i})$, it holds that:
    \begin{equation*}
        p_i(v, f(v)) \leq \eta_i.
    \end{equation*}
\end{definition}

\noindent We further investigate the effects on the operators' payments. 
\begin{lemma}\label{lemma:v>p}
    Under the mechanism $M'=( \mathcal{X},V, F, P)$  with AC, for each operator $i\in \mathcal{I}$, the payments satisfy:
    \begin{align}
        0 \leq p_i(v,f(v)) \leq v_i(f(v))
    \end{align} 
\end{lemma}
\begin{proof}
    Given the reported valuations $v \in V$ and the chosen investment $a = f(v)$, the following holds:
    \begin{align}
        v_{i}(a) - p_i(v,a) & = v_{i}(a) + \sum_{j \neq i} v_{j}(a) - \sum_{j \neq i} v_{j}(f(v_{-i})) \nonumber \\
        &= \sum_{j \in \mathcal{I}} v_{j}(a) - \sum_{j \neq i} v_{j}(f(v_{-i})). \label{eq:u_p}
    \end{align}
    As $\sum_{j \in \mathcal{I}} v_{j}(a)$ represents the maximal social welfare generated by the candidate projects and the environment exhibits no-negative externalities, \eqref{eq:u_p} is non-negative. 
    It is worth noting that the admissibility check does not influence this non-negativity, as it does not alter the social welfare ranking of the projects.
    Finally, in the case where no project is selected ($\bar{X} = \emptyset$), it can be observed that by design $v_i(a) = p_i(v, a) = 0$.
\end{proof}

\noindent We introduce the strategy \textit{Budget-Constrained Truthful Reporting} (BTR), which enforces a constraint on operators' reported valuations:
\begin{equation}
    v_{i}(x) = \max \left(0, \min(b_{i}(x) , \eta_i) \right), \quad \forall x \in \bar{X}, \label{eq:bcr} 
\end{equation}

\noindent Based on Lemma \ref{lemma:v>p}, with BTR, the payments of an operator $i$ will not exceed their valuation for projects and their available budget. Therefore, BTR is an admissible preference profile.

\subsubsection{Properties}
We note that the utility from non-participation is not zero in this setting,
as the selected project can impact the revenue of non-participants. 
We then analyze whether it is still beneficial for operators to participate in the mechanism, given that they may still derive benefits as a non-participant.

\begin{lemma}\label{lemma:non-participation}
    Under the mechanism $M'=( \mathcal{X},V, F, P)$  with AC, for operator $i$, the valuation $v_i \in V_i$ outperforms non-participation, i.e., 
    \[
    u_i(v, a) \geq u_i(v', a'),
    \]
    where $v = (v_i, v_{-i})$, $v' = (\mathbf{0}_{|\bar{X}|}, v_{-i})$, $a=f(v)$ and $a'=f(v')$, 
    if the following conditions hold (referred to as the discrepancy and admissibility conditions, respectively):
    \begin{enumerate}
    \setlength{\itemindent}{-10pt}
        \item \text{if} $b_{i}(x) \geq b_{i}(x') \Rightarrow b_{i}(x) - v_{i}(x) \geq b_{i}(x') - v_{i}(x') \quad \forall x,x' \in \bar{X}$,
        \item $\text{if} ~ \alpha < 1 \Rightarrow ~ p_0(v,a) \leq \alpha c_a$
    \end{enumerate}
\end{lemma}

\begin{proof}
    The utility of non-participation is given by:
    \begin{equation*}
        u_i(v',a') = b_{i}(a').
    \end{equation*}
    Assume $v_i \in \mathcal{V}$ satisfies the discrepancy constraint, then: 
    \begin{equation*}
        b_{i}(a) - b_{i}(a') \geq v_{i}(a) - v_{i}(a').
    \end{equation*}
    Subtracting the two utilities gives:
    \begin{align*}
        u_i(v,a) - u_i(v',a')  
        & = b_{i}(a) - b_{i}(a') - \underset{j \neq i}{\sum} v_{j}(a') + \underset{j \neq i}{\sum} v_{j}(a) \\
        &\geq v_{i}(a) - v_{i}(a') - \underset{j \neq i}{\sum} v_{j}(a') + \underset{j \neq i}{\sum} v_{j}(a) \\
        &= \sum_j v_{j}(a) - \sum_j v_{j}(a').
    \end{align*}
    Due to $\sum_j v_{j}(a)$ being the highest social welfare, it follows that $u_i(v,a) \geq u_i(v',a')$. 
    According to the admissibility condition, the participation of $i$ does not hinder the admissibility of the project with the highest social welfare.
\end{proof}


\noindent Then, 
we investigate the conditions when BTR is a partially admissible dominant strategy, i.e. an admissible strategy that dominates all other admissible strategies.

\begin{definition}
    [Partially Dominant Strategy] 
    A preference profile $v_i \in V^a_i$ is a \emph{partially dominant} strategy if it satisfies the following conditions:
    \begin{enumerate}
        \setlength{\itemindent}{-10pt} 
        \item It is dominant if $v_i \in \text{Int}(V^a_i)$;
        \item It outperforms non-participation if $v_i \in \partial(V^a_i)$.
    \end{enumerate} 
    where $V^a_i$ denotes the set of admissible preference profiles.
\end{definition}

\begin{theorem} \label{therom:dominant}
    With the Subsidized Payment Rule $P$, for each operator $i$, Budget-Constrained Truthful Reporting ($v^{B}_i$) is a partially dominant strategy with $\alpha = 1$.
\end{theorem}

\begin{proof} 
     Based on Lemma \ref{lemma:v>p}, BTR strategy is admissible ($v^{B}_i \in V^a_i$), i.e.,
     \begin{align*}
      p_i(v,a)  \leq v^{B}_{i}(a) \leq \eta_i.
     \end{align*} 
     where $v=(v^{B}_i, v_{-i})$. Next, we prove that the BTR strategy is partially admissible dominant.
     Let $a=f(v)$ and $a'=f(v')$, with $v \neq v'$. 
     Suppose $v^{B}_i \in \text{Int}(V^a_i)$, the utility difference between using BTR and any other strategy for operator $i$ is given by:
     \begin{align*}
         u_i(v,a) - u_i(v',a') &= b_{i}(a) + \underset{j \neq i}{\sum} v_{j}(a) - b_{i}(a') - \underset{j \neq i}{\sum} v_{j}(a') \label{th:delta_u}\\
         &\geq \underset{x \in \bar{X}}{\max} \underset{j}{\sum} v_{j}(x) - b_{i}(a') - \underset{j \neq i}{\sum} v_{j}(a').
     \end{align*}
     The project $a$ is the one that maximizes social welfare, i.e., $\max_{x \in \bar{X}} \sum_j v_j(x)$. With $\alpha = 1$, the project will be implemented.
     Therefore, the maximum social welfare exceeds the other two terms, implying that the BTR strategy is dominant when it is in the interior of the admissible set. 
     Next, consider the case where $v^B_i \in \partial(V^a_i)$, i.e. there exists at least one $ a' \in \bar{X}$ such that $v_{i}(a') = \eta_i$. This implies that $b_{i}(a') \geq \eta_i$. In this case, it is possible for $a'$ to satisfy the following inequalities:
    \begin{align*}
        b_{i}(a') + \underset{j \neq i}{\sum} v_{j}(a') \geq b_{i}(a) + \underset{j \neq i}{\sum} v_{j}(a), \\
        v_{i}(a') + \underset{j \neq i}{\sum} v_{j}(a') \leq b_{i}(a) + \underset{j \neq i}{\sum} v_{j}(a),
    \end{align*}
    which indicates that the strategy is not dominant on the boundary of the admissible set.
    Finally, as the BTR strategy satisfies the discrepancy conditions, by Lemma \ref{lemma:non-participation}, it holds that BTR strategy outperforms non-participation.
\end{proof}

\subsubsection{Discussion on strategic behaviors}
It is important to note that the subsidy provided by the central organization is integrated in a way that preserves the inherent properties of the VCG mechanism. Specifically, even though the central organization covers the difference between the raised payment and the project cost, there is no incentive to underbid. This is because the subsidy does not alter the payment rule for the operators. For the multi-regional network design problem, the proposed mechanism incorperates the operators' budget constraints and the central organization's admissibility check. 
While it is possible that the project that maximizes social welfare is not selected, which requires additional conditions for budget-constrained truthful bidding to be a dominant strategy (in \cref{therom:dominant}).
As a result, operators may report valuations strategically, deviating from truthful bidding.
We propose two potential reporting strategies apart from the BTR.
\paragraph*{Minmax Reporting Strategy}
In this strategy, the operator becomes single-minded, prioritizing only the most beneficial project. 
Specifically, the operator tends to truthfully report their valuations for the project that offers the highest benefit while assigning zero valuation to all other projects and subject to budget constraints.
\begin{align*}
    v_i^x =
    \begin{cases} 
    \max \parens{
    B_i,b_i^x
    }, & \text{if } x = \arg\max\limits_{x \in \bar{X}} b_i^x, \\ 
    0, & \text{otherwise}.
    \end{cases}
\end{align*}
\paragraph*{Proportional Reporting Strategy}
In this strategy, the operator tends to prioritize the higher-valued project for selection while maintaining the preference order. The valuations are adjusted so that the relative differences between valuations remain the same, by dividing by the highest valuation and multiplying by their budget.
\begin{align*}
    v_i^x = \max
    \parens{
    0,
    \frac{b_i^x}{b_i^{x'}}B_i
    }, \quad \text{where } b_i^{x'} = \max\limits_{x \in \bar{X}} b_i^x.
\end{align*}
Those strategies are integrated into the experiments and compared with BTR.
We note that there can be more complex interactions, especially when multiple operators are involved, making the prediction of strategic behaviors non-trivial. 
We present those strategies as the initial step toward exploring bidding strategies and enhancing the mechanism.

\subsection{Baseline}
\noindent The status quo in infrastructure design can be characterized by the local design model (LD), where each operator invests within their own territory and establishes cross-border collaborations with neighboring countries. To ensure a fair comparison with the proposed IVCG mechanism, the available resources for each region should be equal in both cases. Specifically, the subsidy allocated to the region should also be available in LD. 
For each design horizon, the optimization problem for local design can be expressed as:
\begin{align}
\underset{X_i^t \in \mathcal{X}_i}{\text{max}} \, b_i(X_i^t) \quad \text{s.t.} \quad \sum_{a \in X^t_i} c_a \leq B^t_i + B^t_{0i}, \nonumber
\end{align}
where $X_i$ represents the set of projects that can be invested by operator $i$, and $B^t_{0i}$ is the subsidy allocated to operator $i$.
By employing this baseline, we could examine the efficiency of our mechanism in fostering collaboration among operators, as the 
performance differences will stem from how projects are selected, rather than the subsidy amount.


\section{Numerical Experiments} \label{sec:exp}
\noindent 
The numerical experiments are based on a subset of the European Railway Network, including seven countries: Germany, Austria, Luxembourg, Italy, Switzerland, France, and Belgium. 
The primary railway connections are extracted from OpenStreetMap, along with the locations of city centers, the closest train stations, and highway junctions. Airport locations and air travel connectivity data are from OpenFlights.
Based on population statistics, 69 major cities are selected, and travel demand is generated accordingly.

\begin{figure}[tb]
    \centering
    \includegraphics[width=1\linewidth]{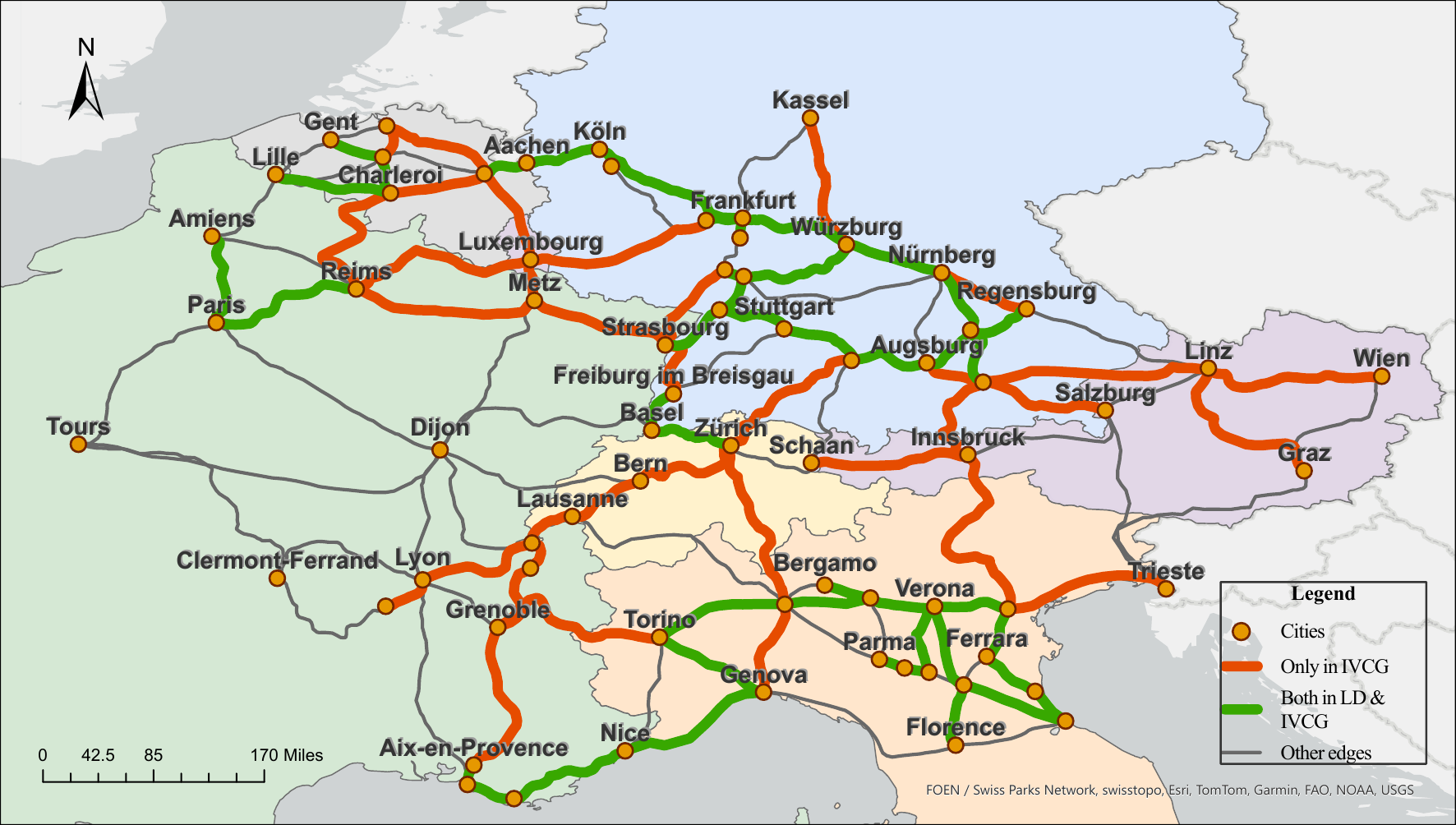}
    \caption{The European railway networks.}
    \label{fig:map}
\end{figure}

\noindent
Given a budget of 32.5 B€ from the central organization with $\alpha = 1$, Figure \ref{fig:map} shows the designed European railway network. 
Red lines indicate implemented connections via the mechanism but not with the local design, whilst green connections are implemented in both cases.
It is evident that the proposed mechanism strategically allocates subsidies to improve cross-border links and reinforce key regional connections.

\begin{figure}[tb]
    \centering
    \includegraphics[width=0.9\linewidth]{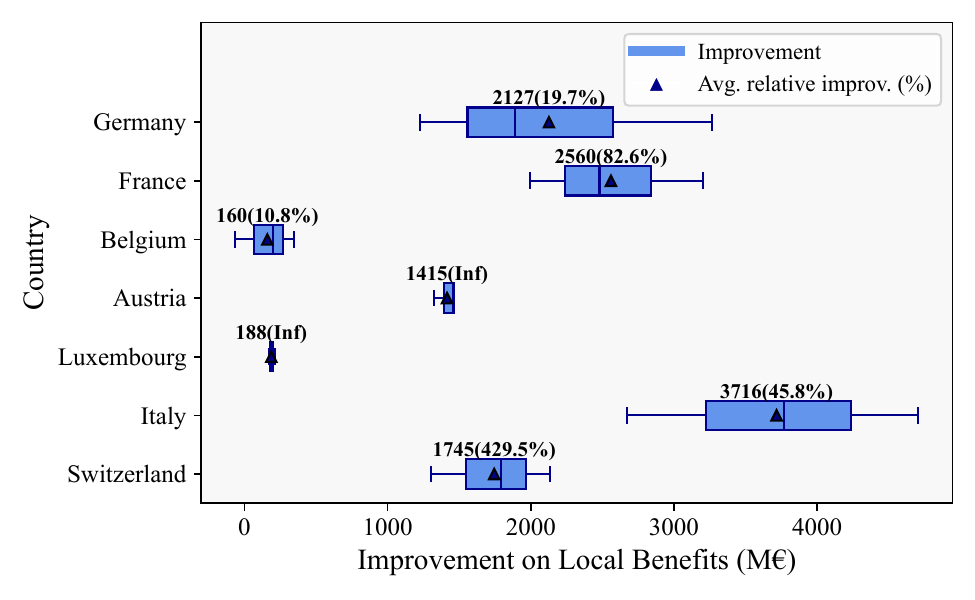}
    \caption{Improvements of the Local Benefits.}
    \label{fig:local}
    \vspace{-5pt}
\end{figure}

\noindent 
Figure \ref{fig:local} presents the comparative improvements in social welfare under the proposed mechanism versus the local design with various investment ratios ($\alpha \in \{0.6, 0.8, 1.0\}$). The overall social welfare is improved by 49.7\% in these scenarios.
The results indicate that, under varying levels of willingness to subsidize, participating countries tend to achieve greater benefits through the proposed IVCG mechanism.
In terms of absolute social welfare improvements, Italy demonstrates the highest average gain, followed by France. 
When considering the relative percentage improvements in social welfare compared to LD, Switzerland has the most substantial increase, with its improvement exceeding four times that of LD. 
This is likely due to its geographic position, being surrounded by neighboring countries, then cross-border projects become more impactful in local social welfare. 

\begin{figure}[tb]
    \centering
    \includegraphics[width=1\linewidth]{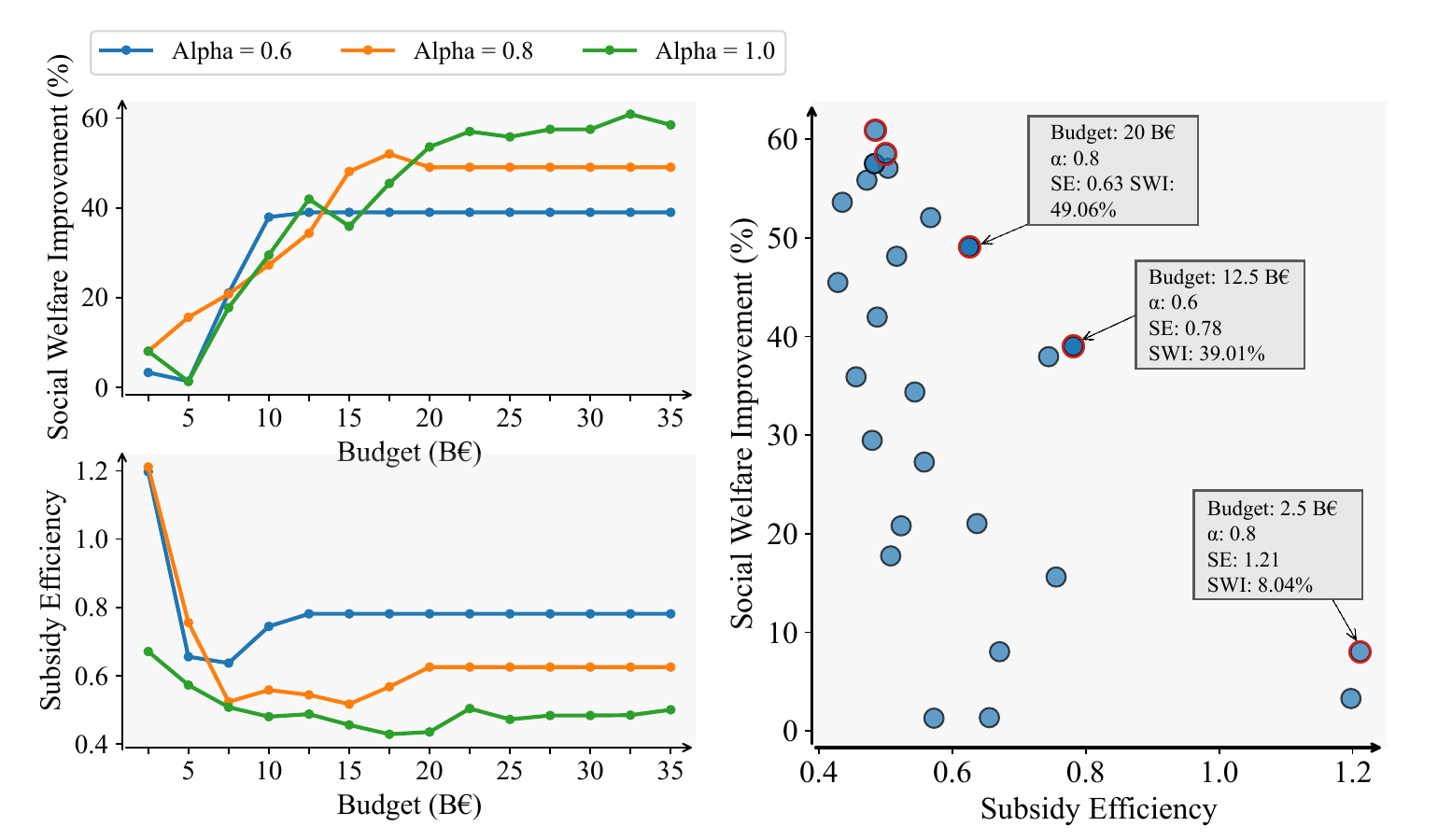}
    \caption{Social welfare improvement and subsidy efficiency.}
    \label{fig:combined}
\end{figure}

\noindent 
The influence of different central budgets and investment ratios on social welfare and subsidy efficiency is shown in Figure \ref{fig:combined}.
An increased budget from the central organization generally enhances the system's social welfare compared to the local design scenario.
Additionally, with a low-level budget, stricter admissibility checks could outperform.
For example, under $B_0$ = 10 B€, an investment ratio of 0.6 results in the highest improvement in social welfare of 40\%.
This suggests that a more conservative approach could be beneficial for the central organization under budget constraints.
In contrast, a higher investment ratio facilitates the optimal utilization of larger budgets.
For instance, when the budget exceeds 20 B€, the social welfare remains unchanged with investment ratios less than 1.

\begin{figure}[htb!]
    \centering
    \includegraphics[width=1\linewidth]{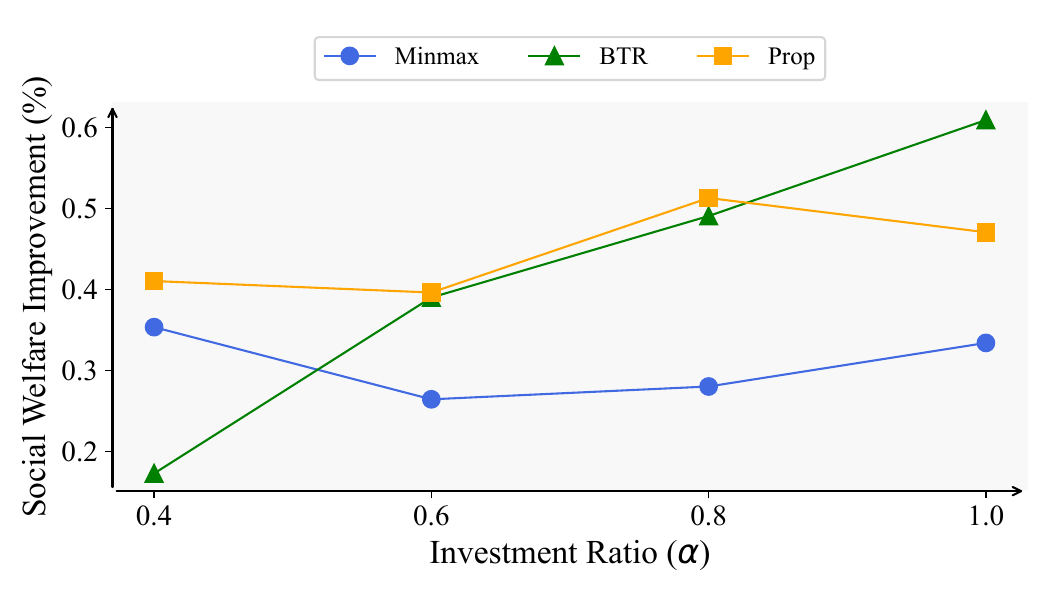}
    \caption{Social welfare under various bidding strategies.}
    \label{fig:str}
\end{figure}

\noindent 
Figure \ref{fig:str} analyzes the impact of different bidding strategies under central investment restrictions ($\alpha =\{0.6,0.8\}$). 
It shows that deviation from the BTR strategy does not necessarily lead to a decline in social welfare. 
In some cases, operators may enhance system-wide outcomes by becoming more pivotal and contributing higher payments. 
However, when the central organization has a higher acceptance of proposals, min-max bidding can negatively affect system performance, as it distorts the true priority order of projects.

\section{Conclusion} \label{sec:conclusion}
Coordination among subsystems is critical to the performance of network-based systems. 
However, cooperative decisions may be infeasible due to insufficient input, even when the interests of self-interested network designers align.
In this work, we explore how an upper-level decision-maker can efficiently allocate subsidies to foster cooperation and enhance overall social welfare.
Specifically, we propose an iterative VCG-based mechanism that incorporates the central organization's preferences and operators' budget constraints, enabling iterative decision-making for multi-year infrastructure investments.
The European railway system experiment demonstrates the efficiency of the proposed mechanism in stimulating cooperation and compares it with local design cases.
Results show that the total budget of the central organization and the expected investment efficiency significantly impact social welfare improvements. 
In addition, it is crucial for the central organization to obtain the true order of the projects from the operators in terms of local social welfare.

\noindent
To improve the current framework, we identify some potential directions for future research.
First, the investment cycle can be improved by considering the central organization’s funding source and the redistribution of collected payments that exceed the actual cost of the project.
In addition, advancing mechanism design can be explored to improve properties (for example incentive compatibility) under budget constraints of operators and the willingness of the central organization to contribute.
Lastly, stable long-term cooperation requires developing robust strategies that consider investment risks, population changes, and the emergence of new transportation modes.


\newpage
\bibliographystyle{IEEEtran}
\bibliography{ref}

@article{small_VoT,
    title = {Valuation of travel time},
    journal = {Economics of Transportation},
    volume = {1},
    number = {1},
    pages = {2-14},
    year = {2012},
    issn = {2212-0122},
    doi = {https://doi.org/10.1016/j.ecotra.2012.09.002},
    url = {https://www.sciencedirect.com/science/article/pii/S2212012212000093},
    author = {Kenneth A. Small}
}

@article{nie2022strategies,
  title={Strategies for applying carbon trading to the new energy vehicle market in China: An improved evolutionary game analysis for the bus industry},
  author={Nie, Qingyun and Zhang, Lihui and Tong, Zihao and Hubacek, Klaus},
  journal={Energy},
  volume={259},
  pages={124904},
  year={2022},
  publisher={Elsevier}
}

@inproceedings{deng2021towards,
  title={Towards efficient auctions in an auto-bidding world},
  author={Deng, Yuan and Mao, Jieming and Mirrokni, Vahab and Zuo, Song},
  booktitle={Proceedings of the Web Conference 2021},
  pages={3965--3973},
  year={2021}
}

@article{ding2023mechanism,
  title={Mechanism design for Mobility-as-a-Service platform considering travelers’ strategic behavior and multidimensional requirements},
  author={Ding, Xiaoshu and Qi, Qi and Jian, Sisi and Yang, Hai},
  journal={Transportation Research Part B: Methodological},
  volume={173},
  pages={1--30},
  year={2023},
  publisher={Elsevier}
}

@inproceedings{zardini2021game,
  title={Game theory to study interactions between mobility stakeholders},
  author={Zardini, Gioele and Lanzetti, Nicolas and Guerrini, Laura and Frazzoli, Emilio and D{\"o}rfler, Florian},
  booktitle={2021 IEEE International Intelligent Transportation Systems Conference (ITSC)},
  pages={2054--2061},
  year={2021},
  organization={IEEE}
}

@article{sessa2017exploring,
  title={Exploring the Vickrey-Clarke-Groves mechanism for electricity markets},
  author={Sessa, Pier Giuseppe and Walton, Neil and Kamgarpour, Maryam},
  journal={IFAC-PapersOnLine},
  volume={50},
  number={1},
  pages={189--194},
  year={2017},
  publisher={Elsevier}
}

@article{karaca2019designing,
  title={Designing coalition-proof reverse auctions over continuous goods},
  author={Karaca, Orcun and Sessa, Pier Giuseppe and Walton, Neil and Kamgarpour, Maryam},
  journal={IEEE Transactions on Automatic Control},
  volume={64},
  number={11},
  pages={4803--4810},
  year={2019},
  publisher={IEEE}
}

@article{ren2024truthful,
  title={Truthful Auction Mechanisms for Dependent Task Offloading in Vehicular Edge Computing},
  author={Ren, Hualing and Liu, Kai and Yan, Guozhi and Liu, Chunhui and Li, Yantao and Li, Chuzhao and Wu, Weiwei},
  journal={IEEE Transactions on Mobile Computing},
  year={2024},
  publisher={IEEE}
}

@misc{iija,
  author       = {{Library of Congress}},
  title        = {{H.R.3684 - 117th Congress (2021-2022): Infrastructure Investment and Jobs Act}},
  year         = {2021},
  month        = {November 15},
  howpublished = {\url{https://www.congress.gov/bill/117th-congress/house-bill/3684}},
}

@misc{bri,
  author       = {{World Bank Group}},
  title        = {{Belt and Road Initiative}},
  year         = {2018},
  url          = {https://www.worldbank.org/en/topic/regional-integration/brief/belt-and-road-initiative},
  note         = {Accessed: 2025-02}  
}

@misc{tarn,
  author       = {{Economic and Socal Commission for Asia and Pacific}},
  title        = {{Trans-Asian Railway Network}},
  url          ={https://www.unescap.org/ourwork/transport/regional-land-transport/trans-asian-railway-network},
  note         = {Accessed: 2025-02}  
}

@misc{TenT,
  author       = {{European Commission }},
  title        = {{Trans-European Transport Network (TEN-T)}},
  year         = {2024},
  url          = {https://transport.ec.europa.eu/transport-themes/infrastructure-and-investment/trans-european-transport-network-ten-t_en},
  note         = {Accessed: 2025-02} 
}

@misc{IEA2023,
  author       = {{International Energy Agency (IEA)}},
  title        = {Tracking Rail},
  year         = {2023},
  month        = {July 11},
  url          = {https://www.iea.org/energy-system/transport/rail#tracking},
  note         = {Accessed: 2025-02}  
}

@misc{railmarket2024,
  author       = {{Market Research}},
  title        = {Rail Infrastructure Market Forecasts to 2030},
  year         = {2024},
  month        = {November},
  url          = {https://www.marketresearch.com/Stratistics-Market-Research-Consulting-v4058/Rail-Infrastructure-Forecasts-Global-Type-38950268/},
  note         = {Accessed: 2025-02}  
}

@article{cats2025long, 
    title={The long journey towards a shift to rail in the European long-distance passenger transport market}, 
    author={Cats, Oded}, 
    journal={npj Sustainable Mobility and Transport}, 
    volume={2}, 
    number={1}, 
    pages={7}, 
    year={2025}, 
    publisher={Nature Publishing Group UK London} }

@misc{eurostat,
  author       = {{Eurostat}},
  title        = {Modal split of air, sea and inland passenger transport},
  year         = {2024},
  month        = {July},
  note         = {Accessed: 2025-02. Available:   \url{https://ec.europa.eu/eurostat/databrowser/view/tran_hv_ms_psmod/default/table?lang=en}} 
}

@inproceedings{he_co-investment_2024,
      title={Co-investment with Payoff Sharing Benefit Operators and Users in Network Design}, 
      author={Mingjia He and Andrea Censi and Emilio Frazzoli and Gioele Zardini},
      booktitle={2025 American Control Conference (ACC)}, 
      year={2025}
}

@inproceedings{zardini2023strategic,
  title={Strategic interactions in multi-modal mobility systems: A game-theoretic perspective},
  author={Zardini, Gioele and Lanzetti, Nicolas and Belgioioso, Giuseppe and Hartnik, Christian and Bolognani, Saverio and D{\"o}rfler, Florian and Frazzoli, Emilio},
  booktitle={2023 IEEE 26th International Conference on Intelligent Transportation Systems (ITSC)},
  pages={5452--5459},
  year={2023},
  organization={IEEE}
}

@article{grolle2024service, 
        title={Service design and frequency setting for the European high-speed rail network}, 
        author={Grolle, Jorik and Donners, Barth and Annema, Jan Anne and Duinkerken, Mark and Cats, Oded}, 
        journal={Transportation Research Part A: Policy and Practice}, 
        volume={179}, 
        pages={103906}, 
        year={2024}, 
        publisher={Elsevier} }

@inproceedings{luo2021multimodal, 
        title={Multimodal mobility systems: joint optimization of transit network design and pricing}, 
        author={Luo, Qi and Samaranayake, Samitha and Banerjee, Siddhartha}, 
        booktitle={Proceedings of the ACM/IEEE 12th International Conference on Cyber-Physical Systems}, 
        pages={121--131}, 
        year={2021} }

@incollection{cramton_lovely_2005,
	title = {The {Lovely} but {Lonely} {Vickrey} {Auction}},
	isbn = {978-0-262-03342-8},
	url = {https://academic.oup.com/mit-press-scholarship-online/book/29406/chapter/244776729},
	language = {en},
	urldate = {2024-11-28},
	booktitle = {Combinatorial {Auctions}},
	publisher = {The MIT Press},
	author = {Ausubel, Lawrence M. and Milgrom, Paul},
	editor = {Cramton, Peter and Shoham, Yoav and Steinberg, Richard},
	month = dec,
	year = {2005},
	doi = {10.7551/mitpress/9780262033428.003.0002},
	pages = {17--40},
}

@book{green_incentives_1978,
	series = {Studies in public economics},
	title = {Incentives in public decision-making},
	isbn = {0-444-85144-5},
	publisher = {North-Holland Pub. Co.},
	author = {Green, Jerry R},
	year = {1978},
}

@article{le2018pareto,
  title={Pareto optimal budgeted combinatorial auctions},
  author={Le, Phuong},
  journal={Theoretical Economics},
  volume={13},
  number={2},
  pages={831--868},
  year={2018},
  publisher={Wiley Online Library}
}

@report{DB2024,
  author       = {{Deutsche Bahn}},
  title        = {Bilanzpressekonferenz 2024},
  year         = {2024},
  institution  = {Deutsche Bahn AG},
  url          = {https://www.deutschebahn.com/de/presse/suche_Medienpakete/Bilanzpressekonferenz-2024-12718954},
  note         = {Accessed: 2025-03-23}
}

@report{SCNF2023,
  author       = {{SNCF Group}},
  title        = {{SNCF Group Annual Results 2023}},
  year         = {2024},
  institution  = {SNCF Group},
  url          = {https://www.groupe-sncf.com/medias-publics/2024-03/presentation-sncf-group-2023-results.pdf},
}

@report{FSGroup2024,
  author       = {{FS Group}},
  title        = {2025–2029 Strategic Plan},
  year         = {2024},
  institution  = {Ferrovie dello Stato Italiane},
  url          = {https://www.fsitaliane.it/content/fsitaliane/en/media/press-releases/2024/12/12/fs-group-2025-2029-strategic-plan.html},
}

@report{Infrabel2023,
  author       = {{Infrabel}},
  title        = {Annual Report 2023},
  year         = {2024},
  institution  = {Infrabel},
  url          = {https://infrabel.be/sites/default/files/generated/files/report/Annual%20Report%202023.pdf},
}

@report{SBB2023,
  author       = {{SBB}},
  title        = {Facts and Figures - Finance},
  year         = {2024},
  institution  = {Swiss Federal Railways (SBB)},
  url          = {https://reporting.sbb.ch/en/finance?highlighted=1fbcf55f8e27556eb094357d9d5b7370&years=1,4,5,6,7&scroll=1836},
  note         = {Accessed: 2025-03-23}
}

@report{OBB2023,
  author       = {{ÖBB-Infrastruktur AG}},
  title        = {Zahlen, Daten, Fakten 2024},
  year         = {2024},
  institution  = {{ÖBB-Infrastruktur AG}},
  url          = {https://infrastruktur.oebb.at/dam/jcr:03a3177a-b9a6-41aa-91cb-43d8e289b11f/zahlen-daten-fakten-folder-2024-oebb.pdf},
}

@online{CFL2024,
  author       = {{RTL Today}},
  title        = {{CFL to invest €7 billion over 15 years in rail infrastructure}},
  year         = {2024},
  url          = {https://today.rtl.lu/news/luxembourg/a/2047688.html},
}

@online{NZZGleisnutzung,
  author       = {{Neue Zürcher Zeitung}},
  title        = {Schienenunterhalt soll besser werden},
  year         = {2011},
  url          = {https://www.nzz.ch/schweiz/schienenunterhalt-soll-besser-werden-ld.94834},
  note         = {Accessed: 2025-03-23}
}

@article{air_passenger_vot,
  author = {Steven Landau and Geoffrey D. Gosling and Kenneth Small and Thomas Adler},
  title ={Measuring Air Carrier Passengers’ Values of Time by Trip Component},
  journal = {Transportation Research Record},
  volume = {2569},
  number = {1},
  pages = {24-31},
  year = {2016},
  doi = {10.3141/2569-03},
}

@article{VOTEurope_car_train,
    author = {Mark Wardman and V. Phani K. Chintakayala and Gerard {de Jong}},
  title = {Values of travel time in Europe: Review and meta-analysis},
  journal = {Transportation Research Part A: Policy and Practice},
  volume = {94},
  pages = {93-111},
  year = {2016},
  issn = {0965-8564},
  doi = {https://doi.org/10.1016/j.tra.2016.08.019},
}

@report{ECA2018,
  author       = {{European Court of Auditors}},
  title        = {{A European high-speed rail network: not a reality but an ineffective patchwork}},
  institution  = {European Court of Auditors},
  year         = {2018},
  number       = {Special Report No 19/2018},
  url          = {https://www.eca.europa.eu/en/Pages/DocItem.aspx?did=47211},
}

@online{MacrotrendsInflation,
  author       = {{Macrotrends}},
  title        = {European Union Inflation Rate 1999–2023},
  year         = {2024},
  url          = {https://www.macrotrends.net/global-metrics/countries/EUU/european-union/inflation-rate-cpi},
  note         = {Accessed: 2025-03-24}
}

@article{transport_cost_beta,
  title = {The paired combinatorial logit model: properties, estimation and application},
  journal = {Transportation Research Part B: Methodological},
  volume = {34},
  number = {2},
  pages = {75-89},
  year = {2000},
  issn = {0191-2615},
  doi = {https://doi.org/10.1016/S0191-2615(99)00012-0},
  author = {Frank S. Koppelman and Chieh-Hua Wen},
}

@online{CINEACEFTransport,
  author       = {{CINEA }},
  title        = {{Transport Infrastructure under the Connecting Europe Facility (CEF)}},
  year         = {2024},
  url          = {https://cinea.ec.europa.eu/programmes/connecting-europe-facility/transport-infrastructure_en},
  note         = {Accessed: 2025-03-24}
}
\begin{appendices}
\section{Model Parameters and Experimental Setup}
\label{appendix:para}
\begin{table}[h]
    \centering
    \caption{Parameters utilized for multi-regional network design model and numerical experiment scenarios.}
    \label{tab:parameters}
    \resizebox{0.98\linewidth}{!}{%
    \begin{tabular}{llll}
        \hline
        \textbf{Variable}&\textbf{Definition} & \textbf{Value} & \textbf{Reference}\\
        \hline
        \multicolumn{4}{l}{\textbf{Travel Demand}}\\
        $\gamma^{R}_t$&VoT train & 29.75 €/h &~\cite{VOTEurope_car_train}\\
        $\gamma^{C}_t$&VoT car & 20.35 €/h &~\cite{VOTEurope_car_train}\\
        $\gamma^{P}_t$&VoT plane & 46.76 €/h &~\cite{air_passenger_vot}\\
        $\gamma_{a}$& VoT access/egress & 17.05 €/h &~\cite{air_passenger_vot}\\
        $\gamma_\mathrm{w}$& VoT waiting time & 25 €/h &~\cite{air_passenger_vot}\\
        $\beta_\mathrm{c}$& travel cost sensitivity & 0.0461 &~\cite{transport_cost_beta}\\
        $v^{u}$& average car speed (urban) & 38 km/h & -\\
        $v^{m}$&average car speed (motorway) & 116.5 km/h & -\\
        $v^{p}$&plane speed & 880 km/h & -\\
        $v^{R}$&train speed & 148 km/h &~\cite{ECA2018}\\
        $t^{P}_\mathrm{w}$&plane waiting time & 1.5 h & - \\
        $p_{h}$&ticket price per hour & 50.4 €/h &~\cite{ECA2018}\\
        \multicolumn{4}{l}{\textbf{Network Design}}\\
        $T$&Planning horizon & 7 year & -\\
        $a_{EU}$&EU budget & 5285 M€ &~\cite{CINEACEFTransport}\\
        $a_{G}$&Germany budget & 2600 M€ &~\cite{DB2024}\\
        $a_{F}$&France budget & 5500M€ &~\cite{SCNF2023}\\
        $a_I$&Italy budget & 5000 M€ &~\cite{FSGroup2024}\\
        $a_L$&Luxembourg budget & 466.7 M€&~\cite{CFL2024}\\
        $a_S$&Switzerland budget & 2210 M€ &~\cite{SBB2023}\\
        $a_B$&Belgium budget & 1100 M€&~\cite{Infrabel2023}\\
        $a_A$&Austria budget & 1750 M€ &~\cite{OBB2023}\\
        $c_{km.}$&cost per km construction & 15.7 M€/km &~\cite{ECA2018}\\
        $T_{R}$& rail lifetime & 33 yrs. &~\cite{NZZGleisnutzung}\\
        $\delta$ & discount factor & 0.976 &~\cite{MacrotrendsInflation}\\ 
        \hline
    \end{tabular}
    }
\end{table}

\end{appendices}
\end{document}